\documentclass[manuscript]{acmart}

\setcopyright{none}

\usepackage{booktabs}
\usepackage{multirow}
\usepackage{graphicx}
\usepackage{fancybox}
\usepackage{caption}
\usepackage{adjustbox}
\usepackage{diagbox}
\usepackage{enumitem}
\usepackage{makecell}
\usepackage{booktabs}
\usepackage{multirow}
\usepackage{diagbox}
\usepackage{graphicx}
\usepackage{array}
\usepackage{xcolor}
\usepackage{colortbl}
\usepackage{graphicx}
\usepackage{subcaption}
\usepackage{fancybox} 

\begin{document}

\title{AWARE Narrator and the Utilization of Large Language Models to Extract Behavioral Insights from Smartphone Sensing Data}

\renewcommand{\shorttitle}{AWARE Narrator}

\author{Tianyi Zhang}
\email{t.zhang59@student.unimelb.edu.au}
\orcid{0000-0002-0778-8844}
\affiliation{%
  \institution{University of Melbourne}
  \city{Melbourne}
  \country{Australia}
}

\author{Miu Kojima}
\email{kojima.m.ap@m.titech.ac.jp}
\orcid{0009-0006-6122-6750}
\affiliation{%
  \institution{Institute of Science Tokyo}
  \city{Tokyo}
  \country{Japan}
}

\author{Simon D'Alfonso}
\email{dalfonso@unimelb.edu.au}
\orcid{0000-0001-7407-8730}
\affiliation{%
  \institution{University of Melbourne}
  \city{Melbourne}
  \country{Australia}
}

\begin{abstract}
Smartphones, equipped with an array of sensors, have become valuable tools for personal sensing. Particularly in digital health, smartphones facilitate the tracking of health-related behaviors and contexts, contributing significantly to digital phenotyping, a process where data from digital interactions is analyzed to infer behaviors and assess mental health. Traditional methods process raw sensor data into information features for statistical and machine learning analyses. In this paper, we introduce a novel approach that systematically converts smartphone-collected data into structured, chronological narratives. The AWARE Narrator translates quantitative smartphone sensing data into English language descriptions, forming comprehensive narratives of an individual's activities. We apply the framework to the data collected from university students over a week, demonstrating the potential of utilizing the narratives to summarize individual behavior, and analyzing psychological states by leveraging large language models.
\end{abstract}


\maketitle

\section{Introduction}

The modern smartphone contains an array of sensors that enable the sensing and tracking of various phone states, uses and properties. These sensors include accelerometer, GPS/geolocation, Bluetooth, communication logs (phone and SMS), application usage and keyboard activity. Given their various sensors and the opportunities to utilise them, smartphones, the Swiss army knives of digital technology, have proven to be valuable personal sensing devices, with applications in domains such as health, education and leisure.

Given their potential to track various health-related behaviours and user contexts, as well as the emergence of health apps, smartphone sensing has become a pivotal topic in digital health. This is particularly the case in digital mental health, where the concept of digital phenotyping has emerged in recent years. In short, digital phenotyping espouses the idea that the data created from our use of and interaction with digital technologies, such as smartphones, can be mined or analysed to infer behaviours and, ultimately assess mental health \cite{baumeister2019digital,bufano2023digital}. The focus of our work in this paper is on leveraging smartphone sensing as a tool in psychology and mental health.

Once raw sensor data is collected, it is typically processed into information features that can be used in statistical analyses and machine learning model construction. For instance, from raw geolocation data one, features such as total distance travelled or time spent at the most visited location can be derived. In this paper, however, we propose a novel approach to analyze smartphone sensing data. The core idea is to translate quantitative smartphone sensing data records into corresponding descriptions in English (or other natural language), which could ultimately be used to construct a narrative summary that describes an individual’s day (or other specified period).

The generation of sensing statements based on data from digital sensors, let alone modern smartphone sensors, has received limited attention in the literature \cite{reddington2011automatically}. Beyond general interest, from our perspective the idea of translating smartphone data records into English descriptions has a further motivation in recent times given the contemporary influence of large language models (LLM) and the availability of systems such as OpenAI’s GPT models and Google’s Gemini. Whilst one approach involves applying LLMs to tabular data for tasks such as description, prediction and general quantitative reasoning \cite{fang2024largelanguagemodelsllmstabular}, another promising avenue is to convert tabular smartphone data records into sets of English statements. These statements could then be fed into LLMs to extract descriptive summaries, generate pattern insights, or even explore the potential for LLMs to make inferences about mental health.

AWARE Narrator systematically and chronologically organizes smartphone-collected data into structured narratives. In this study, we demonstrate the potential of consolidating multi-sensor smartphone data into sensing statements that can be effectively utilized with LLMs. This approach shifts the behavioral analysis problem into the realm of natural language processing. Compared to traditional analysis methodologies (e.g. basic data feature calculations), this approach offers several benefits. First, by integrating data from a wide array of sensors, the sensing statements reveal more information that might not be readily apparent through conventional analysis. Compared to raw quantitative and categorical data, the sensing statements offer finer granularity and richer, human-readable information. Second, this approach provides multi-dimensional information. Each entry point of the sensing statement encapsulates various dimensions such as timeline, sensor type, and detailed target information collected (e.g. message/call contact names or Bluetooth device names). Third, the sensing statements are designed to be interpretable by both humans and machines, thereby enhancing transparency in data analysis. This dual interpretability may better facilitate communication and collaboration between human analysts and automated systems.

\section{AWARE Smartphone Sensing Data}
To illustrate the core ideas of this paper and establish a framework for translating smartphone sensing data into English sensing statements, we utilize the AWARE smartphone sensing platform, in particular, the AWARE-Light app variant \cite{van2023aware}.

The AWARE platform is a versatile and powerful tool designed to harness the sensor capabilities of modern smartphones for both research and practical applications. By integrating a variety of sensors, such as accelerometers, gyroscopes, GPS, and light sensors, AWARE enables the continuous and real-time collection of data related to physical activities, environmental conditions, and user interactions. The data gathered from these sensors is stored in a MySQL database, with each sensor’s data being stored into its own table. This structured data can then be analyzed and interpreted to yield deep insights into user behavior and environmental contexts.

AWARE's open-source nature, along with its robust API and plugin architecture, offers researchers and developers significant flexibility to customize and extend its functionalities to suit specific needs. This adaptability makes AWARE an invaluable resource across a wide range of disciplines, including health monitoring, smart environments, and urban computing, thereby fostering innovation and collaboration within the global research community.

We will now list all the AWARE sensors included in our conversion framework and provide a description of the fields associated with each sensor. It is important to note that, in addition to the specific fields detailed for each sensor, every sensor table includes three standard columns: row id, timestamp, and device ID.

\begin{itemize}[noitemsep, topsep=0pt]
\item Application (foreground) - contains the log of applications the user has interacted with. Each record contains the application name, its Android package name and whether it is an Android system app.
\item Application (notifications) - contains the log of application notifications. As well as application name, this table also records the text content of the notification, although this can be masked if that setting is activated.
\item Battery - contains the battery level of the device
\item Bluetooth - contains the log of nearby scanned devices and connected devices.
\item Calls - contains the log of incoming, outgoing and missing calls, along with masked identifiers
\item Keyboard - contains the input content from the keyboard, which can be masked if set by participants
\item Location - contains the log of longitude and latitude coordinates of the device's location.
\item Messages - contains the log of sent and received messages with masked identifiers
\item Screen - monitors the screen statuses, which can be one of four values: on, off, locked and unlocked.
\item Wifi - contains the log of nearby scanned networks and connected networks.
\end{itemize}

\section{AWARE Narrator: Converting AWARE Tabular Data to Narratives} \label{Narratives}
\begin{figure}
  \centering
  \includegraphics[width=1.1\textwidth]{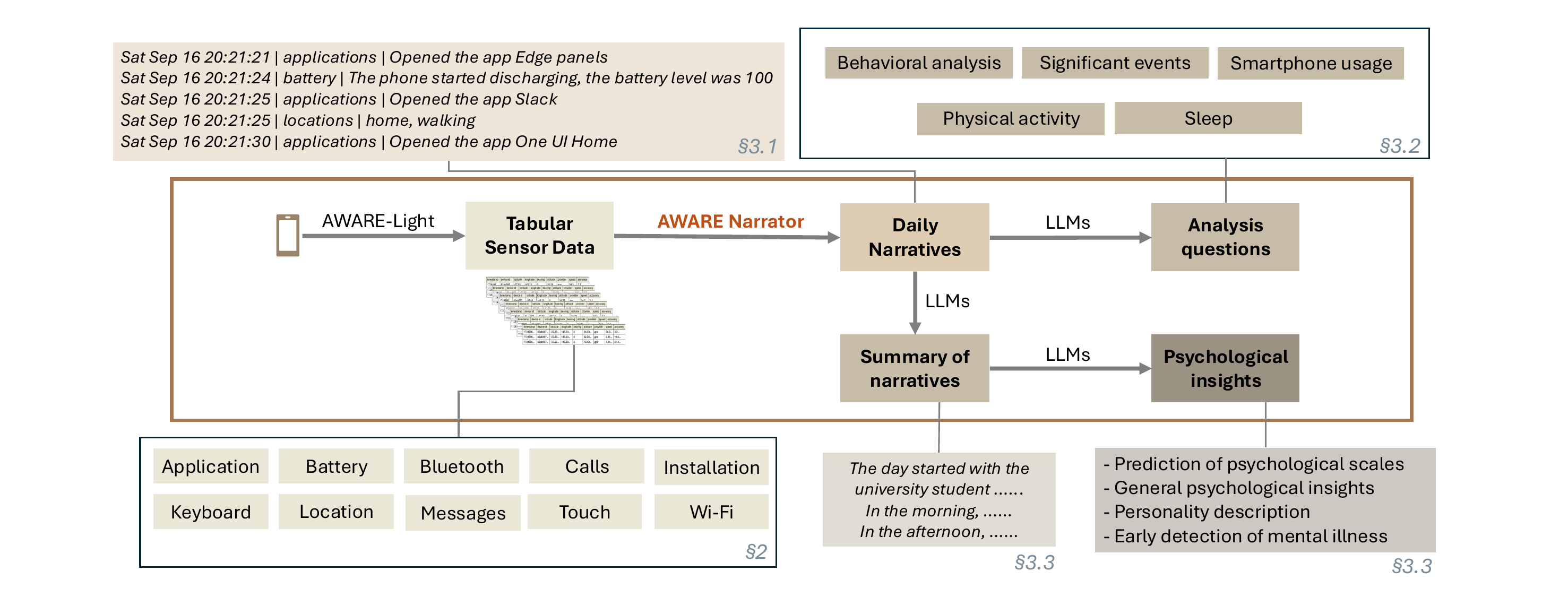}
  \caption{AWARE Narrator workflow.}
  \label{fig:structure}
\end{figure}

In this section, we outline the process of converting tabular data into descriptive sensing statements and utilizing these sensing statements with large language models (LLMs). We begin by collecting smartphone sensor data through AWARE-Light, which is then transformed into daily sensing statements using the AWARE Narrator framework\footnote{\url{https://www.aware-light.org/aware-narrator/}}. These daily sensing statements are subsequently summarized into weekly abstractions to gain psychological insights or analyze behavioral patterns based on daily activities, as depicted in Figure \ref{fig:structure}.

\subsection{Converting AWARE tabular data to descriptive rows}
We have developed a Python script that extracts records from each of the tables mentioned in the previous section, converts each record into an English description, and sorts all the descriptions in chronological order, from earliest to latest. Each line of output is in the following format:

\begin{center}
    <datetime> | <sensor> | <description of sensor record>
\end{center}

A comprehensive list of possible sensor descriptions follows:
\begin{itemize}
\item <datetime> | applications | Opened the app <application name>
\item <datetime> | notifications | Received a notification from the <application name>. The content of the notification was <notification content> (or <datetime> | notifications | Received a notification from the <application name>)
\item <datetime> | battery | <battery status>, the battery level was <battery level>
\item <datetime> | bluetooth | Detected the nearby bluetooth device <bluetooth name>  (or <datetime>| bluetooth | Detected a nearby bluetooth device)
\item <datetime> | bluetooth | Connected to the bluetooth device <bluetooth name> (or <datetime> | bluetooth | Connected to a bluetooth device)
\item <datetime> | calls | <call description> The call lasted <duration> seconds.
\item <datetime> | installations | <application name> was <removed/added/updated>
\item <datetime> | keyboard | Entered the following text into the phone keyboard: <keyboard input>
\item <datetime> | locations | <current place>, <distance>m from home, <stopping/running/walking/riding vehicle> (or | locations | home, <stopping/running/walking/riding vehicle>)
\item <datetime> | messages | Received a message from person <anonymous ID>
\item <datetime> | messages | Sent a message to person <anonymous ID>
\item <datetime> | screen status | Phone screen <turned off/turned on/locked/unlocked>
\item <datetime> | touch | <touch action> <touched content> in the app <application name>
\item <datetime> | wifi | Detected the nearby wifi network <wifi ID> (or <datetime> | wifi | Detected a nearby wifi network)
\item <datetime> | wifi | Connected to the wifi network <wifi ID> (or <datetime> | wifi | Detected a nearby wifi network)
\end{itemize}

The battery status is described as follows:
\begin{itemize}
\item The phone rebooted
\item The phone shutdown
\item The phone started charging
\item The phone started discharging
\item The phone was not charging
\item The phone battery became fully charged
\end{itemize}

The call description is described as follows:
\begin{itemize}
\item Received a phone call from person [anonymous ID]
\item Made a phone call to person [anonymous ID]
\item Missed a call from person [anonymous ID]
\end{itemize}

The touch actions are described as follows:
\begin{itemize}
\item Clicked
\item Clicked longer
\item Scrolled down within a view
\item Scrolled up within a view
\end{itemize}

Note that for Bluetooth and Wi-Fi data, both detected and connected devices are recorded. Detected devices can provide insights into how occupied a location is, while connected devices reflect the user's active connection activity.

For battery data, rather than recording every change, we capture only the local minimum and maximum values. A similar approach is applied to the keyboard sensor, where instead of logging each character typed, we focus on the core content produced by the user, such as complete sentences, individual words, or paragraphs within a session.

Location data is processed using the Density-Based Spatial Clustering of Applications with Noise (DBSCAN) algorithm, specifically employing agglomerative clustering with a 50-meter cluster diameter. The centroid of each cluster is determined using either a manually developed map or the Google Maps API to provide specific location information. Due to the reliance on cluster centroids, this method may result in inaccuracies when retrieving specific location points. Users of the AWARE Narrator framework have the flexibility to adjust the clustering threshold as needed.

The home location is identified based on the cluster containing the most nighttime data points, specifically filtered to include only data between 20:00 and 04:00. Additionally, we calculate the distance from home and determine the corresponding movement status based on recorded speed, defining four types of activity events: stopping (0 m/s), walking (0-1 m/s), running (1-3 m/s), and riding vehicles (over 3 m/s).

For applications running in the foreground, the "System UI" application is excluded as it frequently appears but does not contribute valuable information for analysis. Regarding application notifications, if the content is captured, it is included in the description; otherwise, the description will simply state "Received a notification from <application>." The same approach applies to Bluetooth and Wi-Fi data, where names are omitted if they are not detected.

For messages and calls, AWARE uniquely encrypts each contact's phone number, ensuring privacy while still allowing for the analysis of the number of different contacts and the frequency of communication with each.

\begin{figure}[htbp]
\centering
\fbox{%
    \begin{minipage}{\textwidth}
    \parbox{1\columnwidth}{
        \fontsize{7}{8}\selectfont
        \textsf{The following data is a chronological list that describes the smartphone sensor events collected over a day from the smartphone of a university student. The form of each data record is: timestamp | sensor | description. Answer the following question based on this data: \{question\}.}
        
        \textsf{\{data\}}
    }
    \end{minipage}
}
\caption{Prompt structure of daily questions for descriptive records}
\label{daily_narrative_prompt}
\end{figure}

AWARE Narrator processes sensitive information collected from raw data, such as application notifications, location data, and keyboard inputs. While this private information is preserved and integrated within the AWARE Narrator framework, it can be selectively excluded from the narratives in specific settings to address privacy concerns. This flexibility allows users to balance the richness of the data with the need for privacy protection.

\subsection{Analysis of the sensing statements}

Given a complete set of smartphone sensing data statements for a specific period, we investigate how large language models (LLMs) could be used to summarize the data and extract meaningful insights. For this purpose, we employed the Gemini 1.5-flash, GPT-4o, and Claude Haiku models to perform analysis tasks.

Our first task involves prompting the LLMs with the data and asking them to generate a narrative summary that captures a set of sensing statements for a given day. This summary can serve as a daily report of the individual's activities, which can be valuable for assessing daily events and monitoring mood. When extended over a longer timeline, these narratives can be collected to analyze weekly or monthly changes in behavior and well-being.

We also posed specific questions to the LLMs to extract answers from the sets of sensing statements. These questions can be divided into two categories: factual and analytical. Factual questions focus on descriptive information that can be validated through numeric measures or simple calculations, while analytical questions involve speculation or rely on the LLM's prior knowledge, such as identifying correlational or causal relationships between elements. The questions targeted areas such as smartphone usage, physical activity, sleep, and significant events. Rather than focusing on questions that can be answered through simple calculations, like averages or standard deviations, we emphasized non-numerical, open-ended information that is challenging to infer solely from sensor data. An example is "Are there any patterns in the types of messages (e.g., work-related, social, informational)?" which can vary among different smartphone users. Each question was sent as a separate query to the LLMs. The structure of the prompts is illustrated in Figure \ref{daily_narrative_prompt}. For the reader's benefits, we provide the following list of sample questions that could be applied to such data:

\subsection*{Smartphone Usage}
\begin{itemize}
    \item What are the peak usage times for the person using smartphones throughout the day?
    \item Did the person spend more time interacting with others (e.g. sending messages or making calls), or on their own (e.g. watching videos, playing games, etc.) throughout the day?
    \item What activities does the person use the smartphone for? What kind of role did the smartphone play in the person’s life throughout the day (e.g., working tool, communication tool, or game device; work-related, social, informational)?
    \item Did the user have any representative behavior, such as frequently changing applications, unlocking/locking their phone, checking notifications, or frequently scrolling the application menu?
    \item What is the time distribution of the person using different applications throughout the day?
    \item What actions did the person do in their social media engagement throughout the day (e.g. scrolling, posting posts, or giving likes)?
\end{itemize}

\subsection*{Physical Activity}
\begin{itemize}
    \item How many and what places did the person visit, and what kind of activities did the person presumably conduct at these locations?
    \item What are the frequently visited places for the person?
    \item Was the visited place crowded or not?
    \item How long did the person spend at home?
\end{itemize}

\subsection*{Sleep}
\begin{itemize}
    \item Provide an estimation for how long the person slept.
    \item What patterns emerged regarding the first and last activities before sleep?
\end{itemize}

\subsection*{Significant Events}
\begin{itemize}
    \item Generate a narrative of the day for the person in chronological order.
    \item What are the minor behaviors that you may notice as a large language model that may not be evident or obvious when represented with numeric data?
\end{itemize}

\subsection*{Analysis}
\begin{itemize}
    \item What can be revealed from the provided data (e.g. the keyboard input the person typed, the content they browse and preference of browsed topics)? For example, the personality of the person from their tone when sending messages, or if they initiate or respond to most calls, the schedules/plans of the person, or the opinion/background of the person?
    \item What would be the highlighted events for the person for the day?
    \item What psychological insights into the person can be provided based on their data for the day?
\end{itemize}

\section{Case Study} \label{Case study}
In this section, we apply AWARE Narrator and LLM queries to some data that was obtained from one of the digital phenotyping studies we have run using AWARE-Light, namely StudentSense \cite{dalfonso2024studentsense}. The StudentSense study collected data from university students for 17 weeks during an Australian summer semester at the University of X. Data was collected for the following sensors: application usage, battery, Bluetooth, calls, keyboard usage, geolocation, SMS messages, screen status and WiFi. 

For demonstration purposes, we have selected one of the participants with particularly comprehensive smartphone sensing data and focused on one week of the data during which they were tracked. At the end of the week, psychological measures were administered including the Depression, Anxiety and Stress Scale 21 (DASS-21) \cite{lovibond1995depression} and the International Positive and Negative Affect Schedule (I-PANAS-SF) \citep{thompson2007development}. The sequence of prompts we experimented with to generate summaries and derive insights from the data at both daily and weekly levels are presented.


\begin{figure}[htbp]
\centering
\fbox{%
    \begin{minipage}{\textwidth}
    \parbox{1\columnwidth}{
        \fontsize{7}{8}\selectfont
        \texttt{Thu Sep 14 09:29:10 | applications | Opened the app One UI Home}\par
        \texttt{Thu Sep 14 09:29:11 | applications | Opened the app Phone}\par
        \texttt{Thu Sep 14 09:29:14 | calls | Made a phone call to person 6. The call lasted 0 seconds}\par
        \texttt{Thu Sep 14 09:29:15 | applications | Opened the app Call}\par
        \texttt{Thu Sep 14 09:29:15 | notifications | Received a notification from the Call}\par
        \texttt{Thu Sep 14 09:29:28 | calls | Made a phone call to person 6. The call lasted 0 seconds}\par
        \texttt{Thu Sep 14 09:29:32 | wifi | Connected to the wifi network <unknown ssid>}\par
        \texttt{Thu Sep 14 09:29:32 | wifi | Detected the nearby wifi network ``WiFi-A3E2"}\par
        \texttt{Thu Sep 14 09:29:32 | applications | Opened the app Phone}\par
        \texttt{Thu Sep 14 09:29:46 | calls | Made a phone call to person 6. The call lasted 610 seconds}\par
        \texttt{Thu Sep 14 09:29:47 | applications | Opened the app Call}\par
        \texttt{Thu Sep 14 09:29:47 | notifications | Received a notification from the Call}\par
        \texttt{Thu Sep 14 09:30:12 | wifi | Detected a nearby wifi network}\par
        \texttt{Thu Sep 14 09:31:30 | locations | X Sydney Rd, Coburg VIC 3058, 3099.9m from home, stopping}\par
        \texttt{Thu Sep 14 09:31:59 | locations | X Sydney Rd, Coburg VIC 3058, 3099.9m from home, stopping}
    }
    \end{minipage}
}
\caption{AWARE Narrator example}
\label{description_example}
\end{figure}

We convert the raw data into a narrative using the AWARE Narrator framework. Figure \ref{description_example} presents an example segment of the descriptive data generated by AWARE Narrator from the participant's data, with some critical numbers and street names altered for privacy protection. Compared to traditional tabular sensor data, the narrative provides more straightforward context. For instance, we can observe that at around 9:30 AM, the participant made a 10-minute call after several attempts to reach the same person on X Sydney Road. Additional context is provided by the detected WiFi connections; since only two connections were detected, it suggests that the participant was possibly in a less crowded place. This level of abstraction and contextual information is difficult to obtain from numeric data alone but provides more meaningful insights to both clinicians and the users themselves.

From Figure \ref{description_example}, it is evident that multiple events from different sensors can occur simultaneously, and some information may be repeated over time (e.g., the last two lines of location information). 
This simultaneous occurrence and repetition highlight the richness of the narrative data, which can capture overlapping events from different sensors. The detailed log of the person's day includes every details of them interacting with their smartphone and the information and environment they are exposed to. More in-depth analysis can be developed from this detailed plain descriptive text of behaviors.



\subsection{Querying the Sets of Narrative Sensing Statements}

To briefly demonstrate how answers can be extracted from the narrative information, we provide a few simple examples using GPT-4o, queried from one day's data of the participant. The first query addresses phone usage (Figure \ref{example_smartphone_usage1}), the second query explores the behavior of the smartphone user (Figure \ref{example_smartphone_usage2}), and the third query examines the psychology of the smartphone user (Figure \ref{example_smartphone_usage3}). These three examples illustrate how general questions applied to such data can yield meaningful insights.

\begin{figure}[htbp]
\centering
\fbox{%
    \parbox{1\columnwidth}{
    \fontsize{7}{8}\selectfont
    \textsf{\textbf{\textit{What activities does the person use the smartphone for? What kind of role did the smartphone play in the person’s life throughout the day (e.g., working tool, communication tool, or game device; work-related, social, informational)?}}}\\

    \section*{\textsf{Analysis of Smartphone Usage for the University Student}}

    \textsf{Based on the provided data, we can infer several activities and roles that the smartphone played in the university student's life throughout the day. Here is a detailed analysis:}

    \subsection*{\textsf{Activities the Person Uses the Smartphone For:}}
    \begin{enumerate}
        \item \textsf{\textbf{Charging and Battery Monitoring:}}
        \begin{itemize}
            \item \textsf{The phone was frequently charged, and the battery status was monitored multiple times throughout the day.}
        \end{itemize}

        \item \textsf{\textbf{Screen Usage:}}
        \begin{itemize}
            \item \textsf{The phone screen was turned on and off multiple times, indicating active usage.}
        \end{itemize}

        \item \textsf{\textbf{Wi-Fi Connectivity:}}
        \begin{itemize}
            \item \textsf{The phone connected to various Wi-Fi networks throughout the day, suggesting internet usage.}
        \end{itemize}

        \item \textsf{\textbf{Location Tracking:}}
        \begin{itemize}
            \item \textsf{The phone's location was tracked continuously, showing the student's movements from home to various locations such as the university campus, shops, and other places in Melbourne.}
        \end{itemize}

        \item \textsf{\textbf{Application Usage:}}
        \begin{itemize}
            \item \textsf{The student opened the ``Assistive Ball" app at 03:46:20.}
        \end{itemize}

        \item \textsf{\textbf{Communication:}}
        \begin{itemize}
            \item \textsf{There was a missed call from ``person 2" at 16:40:46.}
        \end{itemize}
    \end{enumerate}

    \subsection*{\textsf{Roles the Smartphone Played in the Person’s Life:}}
    \begin{enumerate}
        \item \textsf{\textbf{Communication Tool:}}
        \begin{itemize}
            \item \textsf{The missed call indicates that the phone was used for communication purposes.}
        \end{itemize}

        \item \textsf{\textbf{Informational Tool:}}
        \begin{itemize}
            \item \textsf{The frequent Wi-Fi connections suggest that the phone was used to access information online, possibly for academic purposes or general browsing.}
        \end{itemize}

        \item \textsf{\textbf{Navigation and Location Tracking:}}
        \begin{itemize}
            \item \textsf{The continuous location updates indicate that the phone was used for navigation or location tracking, helping the student move between different places.}
        \end{itemize}

        \item \textsf{\textbf{Utility Tool:}}
        \begin{itemize}
            \item \textsf{The use of the ``Assistive Ball" app suggests that the phone was used for utility purposes, possibly to assist with accessibility or ease of use.}
        \end{itemize}

        \item \textsf{\textbf{Entertainment or Social Tool:}}
        \begin{itemize}
            \item \textsf{Although not explicitly mentioned, the frequent screen usage and Wi-Fi connections could imply that the phone was also used for social media, entertainment, or other leisure activities.}
        \end{itemize}
    \end{enumerate}

    \subsection*{\textsf{Summary of the Smartphone's Role Throughout the Day:}}
    \begin{itemize}
        \item \textsf{\textbf{Morning to Early Afternoon (00:00 - 12:00):}}
        \begin{itemize}
            \item \textsf{The student was mostly at home, with the phone being charged and used intermittently. The phone connected to Wi-Fi and tracked the student's location at home.}
        \end{itemize}

        \item \textsf{\textbf{Midday to Early Evening (12:00 - 18:00):}}
        \begin{itemize}
            \item \textsf{The student moved from home to various locations, including the university campus and nearby shops. The phone continued to track locations and connect to different Wi-Fi networks.}
        \end{itemize}

        \item \textsf{\textbf{Evening to Night (18:00 - 22:00):}}
        \begin{itemize}
            \item \textsf{The student returned home and continued to use the phone, with frequent Wi-Fi connections and location updates. The phone was charged multiple times, and there was a missed call.}
        \end{itemize}
    \end{itemize}

    \subsection*{\textsf{Conclusion:}}
    \textsf{The smartphone played a multifaceted role in the student's life, serving as a communication tool, informational resource, navigation aid, utility device, and possibly an entertainment or social tool. The data indicates that the student relied on the smartphone for various activities throughout the day, both at home and while moving around the city.}
    }
}
\caption{\textsf{Example I. Smartphone usage question answered by GPT-4o: what are the activities observed from the smartphone and what role the smartphone play?}}
\label{example_smartphone_usage1}
\end{figure}

\begin{figure}[htbp]
\centering
\fbox{%
    \parbox{1\columnwidth}{
    \fontsize{7}{8}\selectfont
\textsf{\textbf{\textit{What user behaviors can be revealed from the provided data (e.g., the keyboard input the person typed, the content they browse, and preference of browsed topics)? For example, the personality of the person from their tone when sending messages or if they initiate or respond to most calls, the schedules/plans of the person, or the opinion/background of the person?)}}}
\section*{\textsf{User Behaviors and Patterns}}

\textsf{Based on the provided data, several user behaviors and patterns can be inferred about the university student:}

\subsection*{\textsf{Home and Movement Patterns}}
\begin{itemize}
    \item \textsf{\textbf{Home Location:}} \textsf{The student spends a significant amount of time at home, as indicated by frequent "home, stopping" entries.}
    \item \textsf{\textbf{Frequent Short Movements:}} \textsf{The student often moves short distances from home, typically around 11.0m to 58.0m, suggesting frequent visits to nearby locations or neighbors.}
    \item \textsf{\textbf{Longer Movements:}} \textsf{Occasionally, the student travels further distances, such as to the main campus (995.0m to 1544.6m from home) and other locations up to 3300.7m from home.}
\end{itemize}

\subsection*{\textsf{Daily Schedule}}
\begin{itemize}
    \item \textsf{\textbf{Late Night Activity:}} \textsf{The student is active late into the night, with movements and phone usage recorded past midnight until early morning (e.g., 00:01:27 to 05:10:55).}
    \item \textsf{\textbf{Morning Activity:}} \textsf{The student starts moving around 07:01:43 and continues to be active throughout the day.}
    \item \textsf{\textbf{Campus Visits:}} \textsf{The student visits the main campus and other university-related locations during typical school hours (11:35:06 to 13:21:29).}
\end{itemize}

\subsection*{\textsf{Phone Usage Patterns}}
\begin{itemize}
    \item \textsf{\textbf{Frequent Screen Unlocks:}} \textsf{The student frequently turns on and off their phone screen, indicating regular phone usage.}
    \item \textsf{\textbf{App Usage:}} \textsf{The student uses specific apps like "Assistive Ball" and "System Launcher," suggesting frequent customization of the phone or access to system settings.}
    \item \textsf{\textbf{WiFi Connectivity:}} \textsf{The student connects to various WiFi networks throughout the day, indicating movement between different locations with available WiFi.}
\end{itemize}

\subsection*{\textsf{Charging Habits}}
\begin{itemize}
    \item \textsf{\textbf{Regular Charging:}} \textsf{The phone is frequently charged, often reaching 100\% battery, indicating the student is diligent about keeping their phone charged.}
    \item \textsf{\textbf{Charging at Home:}} \textsf{Most charging events occur at home, suggesting the student ensures their phone is charged before leaving.}
\end{itemize}

\subsection*{\textsf{Social and Academic Life}}
\begin{itemize}
    \item \textsf{\textbf{Campus and Nearby Locations:}} \textsf{Frequent visits to the main campus and nearby streets suggest the student is actively attending classes or university-related activities.}
    \item \textsf{\textbf{Social Interactions:}} \textsf{Movements to various street numbers and establishments indicate the student might be meeting friends or engaging in social activities.}
\end{itemize}

\subsection*{\textsf{Personality and Preferences}}
\begin{itemize}
    \item \textsf{\textbf{Night Owl:}} \textsf{The late-night activity suggests the student might be a night owl, preferring to stay up late and possibly study or socialize during these hours.}
    \item \textsf{\textbf{Tech-Savvy:}} \textsf{Frequent use of system-related apps and settings indicates the student might be tech-savvy or interested in customizing their phone.}
\end{itemize}

\subsection*{\textsf{Potential Study or Work Locations}}
\begin{itemize}
    \item \textsf{\textbf{Frequent Locations:}} \textsf{The student frequently visits specific locations like "street\_number202, 202 A'Beckett St" and "main campus," which could be study spots, work locations, or places of interest.}
\end{itemize}

\subsection*{\textsf{Physical Activity}}
\begin{itemize}
    \item \textsf{\textbf{Walking and Running:}} \textsf{Occasional entries of "walking" and "running" suggest the student engages in physical activity, possibly walking or running to different locations.}
\end{itemize}

\subsection*{\textsf{Connectivity and Network Usage}}
\begin{itemize}
    \item \textsf{\textbf{Multiple WiFi Networks:}} \textsf{The student connects to various WiFi networks, indicating movement between different locations with available internet access, possibly for academic or social purposes.}
\end{itemize}

\section*{\textsf{Conclusion}}
\textsf{The data reveals that the student has a dynamic daily routine with frequent short movements around their home and longer trips to the university campus. They are active late at night and early in the morning, suggesting a flexible or non-traditional schedule. The student is tech-savvy, regularly uses their phone, and ensures it is well-charged. Their movements indicate a balance between academic responsibilities and social activities.}

}}
\caption{\textsf{Example II. Smartphone-inferred information provided by GPT-4o: What behaviors can be inferred from the provided data?}} 
\label{example_smartphone_usage2}
\end{figure}

\begin{figure}[htbp]
\centering
\fbox{%
    \parbox{1\columnwidth}{
    \fontsize{7}{8}\selectfont

\textsf{\textbf{\textit{What psychological insights into the person can be provided based on their data for the day?~\\}}}
    
\section*{\textsf{Psychological Insights from Smartphone Data}}

\textsf{Based on the provided data, several psychological insights can be inferred about the university student's behavior and lifestyle:}

\begin{enumerate}
    \item \textsf{\textbf{Nocturnal Activity:}}
    \begin{itemize}
        \item \textsf{The student was active late into the night and early morning hours, as indicated by phone usage and location changes between 00:14 and 04:03. This suggests a possible nocturnal lifestyle or irregular sleep patterns, which could be due to studying, social activities, or other personal habits.}
    \end{itemize}

    \item \textsf{\textbf{Frequent Movement:}}
    \begin{itemize}
        \item \textsf{The student moved around various locations in Melbourne throughout the day, including multiple stops at different street numbers and establishments. This indicates a high level of mobility and possibly a busy schedule involving various activities or social engagements.}
    \end{itemize}

    \item \textsf{\textbf{Battery Management:}}
    \begin{itemize}
        \item \textsf{The phone's battery level was critically low at 0\% around 13:21, leading to a shutdown. The student then charged the phone intermittently throughout the evening, with frequent disconnections and reconnections to the charger. This behavior might indicate a lack of planning or awareness regarding battery management, or it could be due to the student being in different locations without consistent access to charging facilities.}
    \end{itemize}

    \item \textsf{\textbf{Social and Public Spaces:}}
    \begin{itemize}
        \item \textsf{The student spent significant time in public and social spaces such as Riverfront Plaza, Unity Plaza, and various streets in Melbourne. This suggests a preference for being in social environments or public areas, which could indicate extroverted tendencies or a need for social interaction.}
    \end{itemize}

    \item \textsf{\textbf{Frequent Phone Screen Activity:}}
    \begin{itemize}
        \item \textsf{The phone screen was turned on and off frequently, especially in the evening. This could indicate a high level of phone usage, possibly for communication, social media, or other applications. The frequent screen activity might also suggest a habit of regularly checking the phone, which could be linked to anxiety or a need for constant connectivity.}
    \end{itemize}

    \item \textsf{\textbf{Irregular Charging Patterns:}}
    \begin{itemize}
        \item \textsf{The phone was charged in short bursts, often disconnecting and reconnecting within minutes. This irregular charging pattern might indicate a lack of stable access to power sources or a habit of charging the phone only when necessary.}
    \end{itemize}

    \item \textsf{\textbf{Use of Assistive Apps:}}
    \begin{itemize}
        \item \textsf{The student opened the app ``Assistive Ball" at 02:47, which is often used for accessibility purposes or to enhance phone navigation. This could suggest that the student has specific needs or preferences for phone usage that require assistive technology.}
    \end{itemize}

    \item \textsf{\textbf{Connectivity to Unknown Wi-Fi Networks:}}
    \begin{itemize}
        \item \textsf{The phone connected to various unknown Wi-Fi networks throughout the day. This could indicate that the student frequently moves between different locations with available Wi-Fi, possibly indicating a need for internet access for academic or personal purposes.}
    \end{itemize}

    \item \textsf{\textbf{Physical Activity:}}
    \begin{itemize}
        \item \textsf{The data shows instances of walking, running, and riding a vehicle. This variety in movement types suggests that the student engages in different forms of physical activity, which could be part of their daily routine or related to their transportation methods.}
    \end{itemize}

    \item \textsf{\textbf{Potential Stress or Anxiety:}}
    \begin{itemize}
        \item \textsf{The frequent phone screen activity, irregular sleep patterns, and constant movement might indicate underlying stress or anxiety. The student's behavior could be a response to academic pressures, social obligations, or personal issues.}
    \end{itemize}
\end{enumerate}

\section*{\textsf{Overall Conclusion}}

\textsf{The data suggests that the student leads a dynamic and possibly hectic lifestyle characterized by irregular sleep patterns, frequent movement, and high phone usage. These behaviors could be influenced by academic responsibilities, social activities, or personal habits. Understanding these patterns can help in providing better support and resources for managing their time, stress, and overall well-being.}

}}
\caption{\textsf{Example III. One of the analysis question answered by GPT-4o: What psychological insights can be inferred?}}
\label{example_smartphone_usage3}
\end{figure}

Example I (Figure \ref{example_smartphone_usage1}) explores the relationship between the participant and the smartphone. This example can serve as a summary for smartphone users, offering daily or periodic reports on their smartphone usage and recommendations for healthier usage habits. Leveraging the LLMs, the smartphone usage data can reveal problematic behaviors such as excessive use, smartphone addiction, and nomophobia (``NoMobilePhobia"). In addition, the example captures and classifies application usage, a task that would be challenging to perform manually with traditional analysis methods. Translating the tabular form of application usage into textual descriptions makes it clearer for LLMs to analyze, with application names and functionalities. Such a summary of application usage provides comprehensive insights into the user's smartphone behaviors and how they interact with various information sources.

Example II (Figure \ref{example_smartphone_usage2}) demonstrates how the collected smartphone sensor data can be used in narrative form to extract behavioural insights. Specifically, using evidence from various perspectives such as movement patterns, application usage, communication patterns, activity patterns, and connectivity via WiFi and networks, the large language models speculate on the personality and behavior based on descriptive data. We confirm the validity of information including frequent visits, used applications, and activity timelines. Although the analysis does not speculate on personality traits like the Big Five Personality with the open-ended question asked, it can reveal how daily smartphone interactions may reflect personality and psychological states. This understanding can help individuals comprehend how their smartphone usage impacts their psychological well-being.

Example III (Figure \ref{example_smartphone_usage3}) presents the output from LLMs analyzing psychological insights into the participant. The summary suggests that the participant had a ``highly connected, academically focused" routine and ``relied heavily on technology with a structured yet flexible routine." This conclusion is based on the participant's academic activity engagement, social interactions, and location data from their smartphone. The LLMs also observe minor details, such as staying up late, time management habits, and constant app switching. These insights, which are difficult to obtain from raw numeric data, are valuable for analyzing psychological well-being and mental health. Previous studies indicate that psychological states can be reflected in smartphone usage patterns, such as anxiety shown by frequent app switching or depression indicated by keystroke patterns. The AWARE Narrator framework captures and analyzes these minor behavioral details more effectively than traditional feature generation methods.

\subsection{Applying LLMs to the Narrative Data for the Prediction of Psychometric Results}

Regarding the psychological insights question, whether the information provided by the LLM ultimately provides truthful/valid responses is an open question beyond the scope of this paper and would require validation in terms of work that establishes associations between digital footprint data and psychological states. Furthermore, how such psychological summaries could be used by relevant practitioners remains another interesting consideration.

Beyond applying a general query to a phone sensing dataset, our dataset was collected during a study in which psychometric assessments were also collected, including the DASS-21 and I-PANAS-SF as specified earlier. The combination of this weekly sensing data and accompanying psychometric results provides an opportunity to experiment with asking LLMs to predict what questionnaire responses the smartphone user participant provided.

To test this idea, we started by taking the original narrative lists for each of the seven days in the given week and asked the LLM to generate a narrative summary of the day in chronological order. We then combined these seven daily narrative summaries together in order and queried this weekly summary information in terms of questions related to the I-PANAS-SF and DASS-21. The prompt for DASS-21 and I-PANAS-SF are shown in Figure \ref{prompt_DASS} and \ref{prompt_PANAS} respectively:

\begin{figure}[htbp]
\centering
\fbox{%
    \begin{minipage}{.65\textwidth}
    \parbox{1\columnwidth}{
        \fontsize{7}{8}\selectfont
        \textsf{Using the narrative of activities collected from smartphone sensors over the past week, estimate the individual's mental health based on the DASS-21 scale. This scale evaluates three subscales: depression, anxiety, and stress, each with a maximum score of 21. For each subscale, categorize the result into one of the following ranges:} ~\\
        \textsf{\textbf{Normal: 0 to 4}}  \\
        \textsf{\textbf{Mild: 5 to 6}} \\
        \textsf{\textbf{Moderate: 7 to 10}} \\
        \textsf{\textbf{Severe: 11 to 13}} \\
        \textsf{\textbf{Extremely Severe: 14 and above}} ~\\
        
        \textsf{Format your output as follows:} \\
        \textsf{\textbf{Depression: <extent>}} \\
        \textsf{\textbf{Anxiety: <extent>}} \\
        \textsf{\textbf{Stress: <extent>}} \\
        
        \textsf{Narratives data:} \{\textsf{narrative data}\}
        }
        \end{minipage}
    }

\caption{\textsf{Prompt for DASS-21 prediction}}
\label{prompt_DASS}
\end{figure}

\begin{figure}[htbp]
\centering
\fbox{%
    \begin{minipage}{.65\textwidth}
    \parbox{1\columnwidth}{
        \fontsize{7}{8}\selectfont
        \textsf{Given a week's narrative of activity collected from smartphone sensors, estimate the scores for the following affective states} \textsf{that the person would report at the end of the week using the I-PANAS-SF scale.} \textsf{The scale ranges from 1 to 5, with 1 representing 'Never' and 5 representing 'Always':}\par 
        ~\\
        \textsf{Upset}\par
        \textsf{Hostile}\par
        \textsf{Alert}\par
        \textsf{Ashamed}\par
        \textsf{Inspired}\par
        \textsf{Nervous}\par
        \textsf{Determined}\par
        \textsf{Attentive}\par
        \textsf{Afraid}\par
        \textsf{Active}\par
        ~\\
        \textsf{Narratives: \{Narratives for each day of the week\}}
    }
    \end{minipage}
}
\caption{\textsf{Prompt for weekly I-PANAS-SF prediction}}
\label{prompt_PANAS}
\end{figure}

We experimented with three LLMs (GPT-4o, Gemini 1.5 Pro, and Claude-Haiku) to summarize the descriptions generated by the AWARE Narrator for a weekly psychological report. These LLMs were also tasked with predicting scores for the DASS-21 and I-PANAS-SF based on the daily narratives produced by the AWARE Narrator. The results, presented in Table \ref{DASS_results}, and Table \ref{PANAS_results} demonstrate the LLMs' ability to summarize narratives and predict psychological states. We observe that LLMs perform relatively well on I-PANAS-IF tasks and acceptably on DASS-21 tasks. However, due to the limited data and the fact that this is just a demonstration where our focus is not on prediction optimization, accuracy metric evaluations are not provided in this paper. Additionally, prompt engineering was not optimized, nor were additional examples provided to LLMs to enhance prediction performance. This demonstration illustrates how transforming raw tabular data into narratives using the AWARE Narrator can be beneficial in ubiquitous computing. In comparison to feeding raw numerical data into LLMs, which can exceed limits and reduce the information extracted by the models, narratives offer a clearer and more understandable description. These narratives can be more easily comprehended by humans, facilitating better interpretation and analysis.

Previous work has explored similar approaches to predicting psychological measures. For instance, \cite{zhang2024leveragingllmspredictaffective} leveraged LLMs to predict affective states by extracting daily features using RAPIDS, which calculated high-level abstracted features such as the total duration of application usage, the total number of missed calls, and the time spent at home. In contrast, our study directly converts each line of raw data into descriptive English narratives. This approach ensures data completeness, preserving the information of each individual event. The comprehensive narratives produced by the AWARE Narrator can potentially be used in conjunction with LLMs or human inspection for clinical reports, assisting psychiatrists and psychologists in helping individuals with mental health issues. By maintaining detailed event information, our method provides a more thorough basis for psychological assessment and intervention.

\begin{table}[h]
\centering
\caption{Comparison of the Prediction Results of DASS-21 by Different LLMs and Actual Results.}
\resizebox{\textwidth}{!}{
\begin{tabular}{lccccccccccc}
\toprule
Narrative LLM & \multicolumn{3}{c}{Gemini} & \multicolumn{3}{c}{OpenAI} & \multicolumn{3}{c}{Claude} & \multirow{2}{*}{\makecell{Actual \\Results}}  \\
 \cmidrule{2-10}
Prediction LLM & OpenAI & Gemini & Claude & OpenAI & Gemini & Claude & OpenAI & Gemini & Claude & \\
\midrule
Depression & Moderate & Mild & Mild & Moderate & Moderate & Mild & Moderate & Mild & Mild & Normal \\
Anxiety & Moderate & Moderate & Moderate & Moderate & \textbf{Mild} & Moderate & Moderate & Moderate & Normal & Mild \\
Stress & Moderate & Normal & Normal & Moderate & Moderate & Normal & Moderate & Moderate & Moderate & Mild \\
\bottomrule
\end{tabular}
}
\label{DASS_results}
\end{table}

\begin{table}[h]
\centering
\caption{Comparison of the Prediction results of I-PANAS-IF by Different LLMs and Actual Results.}
\resizebox{\textwidth}{!}{
\begin{tabular}{lcccccccccc}
\toprule
Narrative LLM & \multicolumn{3}{c}{Gemini} & \multicolumn{3}{c}{OpenAI} & \multicolumn{3}{c}{Claude} & \multirow{2}{*}{\makecell{Actual \\Results}} \\
\cmidrule{2-10}
Prediction LLM  & OpenAI & Gemini & Claude & OpenAI & Gemini & Claude & OpenAI & Gemini & Claude & \\
\midrule
active & \textbf{5} & 3 & 4 & 3 & \textbf{5} & \textbf{5} & \textbf{5} & 3 & \textbf{5} & 5 \\
determined & \textbf{4} & \textbf{4} & \textbf{4} & 3 & \textbf{4} & \textbf{4} & \textbf{4} & 3 & \textbf{4} & 4 \\
attentive & \textbf{4} & 3 & \textbf{4} & 3 & 3 & \textbf{4} & \textbf{4} & 3 & \textbf{4} & 4 \\
inspired & 3 & 2 & 3 & 2 & 3 & \textbf{4} & 3 & 2 & \textbf{4} & 4 \\
alert & 4 & \textbf{3} & 4 & \textbf{3} & 4 & 4 & 4 & \textbf{3} & 4 & 3 \\
upset & \textbf{2} & 3 & \textbf{2} & \textbf{2} & \textbf{2} & \textbf{2} & \textbf{2} & \textbf{2} & \textbf{2} & 2 \\
hostile & 1 & \textbf{2} & \textbf{2} & 1 & 1 & 1 & 1 & 1 & 1 & 2 \\
ashamed & \textbf{1} & 2 & 2 & \textbf{1} & \textbf{1} & \textbf{1} & \textbf{1} & \textbf{1} & \textbf{1} & 1 \\
nervous & \textbf{2} & \textbf{2} & 3 & \textbf{2} & 3 & \textbf{2} & \textbf{2} & \textbf{2} & \textbf{2} & 2 \\
afraid & \textbf{1} & 2 & 2 & \textbf{1} & \textbf{1} & \textbf{1} & \textbf{1} & \textbf{1} & \textbf{1} & 1 \\
\bottomrule
\end{tabular}
}
\label{PANAS_results}
\end{table}

\section{Future Work and Conclusion}

Our exploratory demonstrations with the LLMs were confined to zero-shot prompts. Future work on this idea should explore multi-shot prompting. One approach would be to include some psychological measures of interest as output for each shot input (i.e., set of sensor descriptions). For example, the set of sensor descriptions for each of several consecutive days could be provided as example inputs, accompanied by psychological assessments taken at the end of those days as example outputs to improve the accuracy of the predictions. Additionally, employing a Retrieval Augmented Generation (RAG) could enhance how contextual information is integrated into the model.

The tasks for the LLM could then include predicting psychological assessment values for subsequent inputs without known outputs. This paradigm may assist in the early detection of mental health issues and mood monitoring. Fine-tuning models can also be developed for personalization or for specific cohorts (e.g., individuals with depressive symptoms) to infer physical and mental health and well-being for self-regulation.

Another line of future work concerns extending the AWARE Narrator tool in terms of the range of descriptions it can generate. This could be done by incorporating other sensors or sources. For example, AWARE-Light includes a novel sensor called the screen reader, which collects text displayed on the screen \cite{teng2024predicting}. This sensor can extract core information and keywords, which can be added to the AWARE Narrator for enhanced narrative content. It could also be done by generating descriptions that combine two or more sensors. For example, ``the user was using the Spotify app whilst travelling on a train", which involves both the application sensor and the movement sensors. 

Given the highly sensitive nature of the data collected from smartphone sensors, privacy concerns can be mitigated by running the AWARE Narrator and LLM analysis system directly on-device, thus transforming fine-grained sensor data into an abstract summary of activities without leaving the phone.

Our investigation has demonstrated the feasibility and usability of the AWARE Narrator framework. Assisted by large language models (LLMs), the generated narrative descriptions can be analyzed to infer behaviours, contexts, psychological characteristics and mental health. This makes the AWARE Narrator a practical and useful tool not only for visualizing and representing one's engagement and behaviors with smartphones but also for understanding the information and environments to which the smartphone user is exposed. It provides a novel way to interpret classical tabular data in digital phenotyping, offering insights into human behaviors and psychological states.

\bibliographystyle{ACM-Reference-Format}
\bibliography{references.bib}

\end{document}